\documentclass[sigconf]{acmart}

\usepackage{subcaption}
\usepackage{graphicx}
\usepackage{array}
\usepackage{multirow}
\usepackage{tabularx}

\AtBeginDocument{
    \providecommand\BibTeX{{
        \normalfont B\kern-0.5em{\scshape i\kern-0.25em b}\kern-0.8em\TeX}
    }
}

\copyrightyear{2019}
\acmYear{2019}
\setcopyright{acmlicensed}

\acmConference[WI '19]{WI '19: ACM International Conference on Web Intelligence}{October 14-17, 2019}{Thessaloniki, Greece}
\acmPrice{15.00}
\acmDOI{10.1145/1122445.1122456}
\acmISBN{978-1-4503-9999-9/18/06}

\begin{document}


\title[Check-It: A plugin for Detecting Fake News and Misinformation on the Web]{Check-It: A plugin for Detecting and Reducing the Spread of Fake News and Misinformation on the Web}

\author{Demetris Paschalides, Chrysovalantis Christodoulou, Rafael Andreou, George Pallis, Marios D. Dikaiakos}
\email{ { dpasch01, cchris47, randre07, gpallis, mdd } @cs.ucy.ac.cy}
\affiliation{
    \institution{Computer Science Department, University of Cyprus}
    \city{Nicosia}
    \country{Cyprus}
}

\author{Alexandros Kornilakis, Evangelos Markatos}
\email{ { kornilak, markatos } @ics.forth.gr}
\affiliation{
    \institution{Department of Computer Science, University of Crete}
    \city{Heraklion, Crete}
    \country{Greece}
}

\renewcommand{\shortauthors}{Paschalides et al.}

\begin{abstract}
Over the past few years, we have been witnessing the rise of misinformation on the Web. People fall victims of fake news during their daily lives and assist their further propagation knowingly and inadvertently. There have been many initiatives that are trying to mitigate the damage caused by fake news, focusing on signals from either domain flag-lists, online social networks or artificial intelligence. In this work, we present Check-It, a system that combines, in an intelligent way, a variety of signals into a pipeline for fake news identification. Check-It is developed as a web browser plugin with the objective of efficient and timely fake news detection, respecting the user's privacy. Experimental results show that Check-It is able to outperform the state-of-the-art methods. On a dataset, consisting of 9 millions of articles labeled as fake and real, Check-It obtains classification accuracies that exceed 99\%.
\end{abstract}



\maketitle

\section{Introduction}

\textbf{Motivation and Problem:} Misinformation is not a recent issue. As early as 1925, when news offices started to connect to each other via wire, the authenticity of information became a concern. Editors did not really know whether the news coming in through the wire was true or not. They could try to infer authenticity based on the source of the news, but still the concern remained: is this piece of news that just came over the wire true or not? Although the concern was there, the editors usually managed to find ways to mitigate it and reduce the intentional misinformation to the minimum possible: after all, the amount of news that came over the wire and could potentially be misinformation was not that large. Unfortunately, the "tsunami" of social media engagement that has swept our lives over the past decade practically exploded the proliferation of misinformation including the associated distribution of fake news \cite{Fernandez:2018}.

\textbf{State of the Art and its Limitations:}  Despite the increasing interest in analyzing fake news in the Web and the development of tools to deal with fake news that had been previously identified, there has been very little work in automatic fake news detection tools. Currently people do not have the tools they need in order to filter out information they are not interested in. For example, if their friends share fake news from time to time, they do not have any way to tell the social media platform "I do not want the fake news my friends (probably) inadvertently propagate. Can you filter the fake news (not my friends!) out of my social feed? Or better yet, can you label the fake news as such? I will then do the filtering out." The main problem stems from the fact that it is difficult to develop classification algorithms to capture fake news. Researchers in \cite{McCreadie2015} studied the feasibility of using a crowdsourcing platform to identify rumours and fake news in social media. According to their research outcomes, the annotators achieve high inter-annotator agreement. In \cite{Zhao2015}, authors found that fake news posts in social media are usually provoking posts (i.e., tweets) from users who raise questions about these posts. In this direction, another approach that has been proposed is the development of browser plugins, such as the B.S. Detector\footnote{~\url{http://bsdetecor.tech}} and the FakerFact\footnote{~\url{https://www.fakerfact.org/}}, which flag content from fake news sources using a constantly-updated list of known fake news sites as a reference point.

\textbf{Approach and Contribution:} The focus of our work is the detection of news content that is fabricated and can be verified to be false. In this paper, we present a plugin that fights disinformation using an automated approach. Our approach is inspired by the way we fight SPAM email messages. Indeed, to fight SPAM, computer scientists have developed SPAM filters: automated programs that scan all email messages of each user, categorize them as SPAM (trash email) or HAM (regular email) and filter the SPAM out of the users mailboxes. In this paper, we follow the exact same approach: we process all information (e.g. tweets, posts, web documents, etc.) that users see online and characterize them as misinformation or not. If we find misinformation we clearly label it so that the user will be warned that he should be careful before believing this current piece of news. For experimental studies, we have developed our system as a plug in for the popular web browsers, namely Google's Chrome and Mozilla's Firefox. However, our method is general and applicable to any browser. A key difficulty in our approach is to combine in an intelligent way a variety of signals in order to decide whether a piece of news is misinformation. Such signals include: the reputation of the person (account) posting the news, the reputation of the web site where the news is hosted, natural language processing features that characterize a fake news article etc. Using a deep learning approach, we combine all these features towards providing a rating that is timely and accurate. Another key aspect of our system is that it protects  the privacy of user (GDPR compliant) since the plugin works locally on the user's browser without the need of external communication. We empirically evaluate our proposed method via extensive experiments on real-world datasets from Twitter and news articles, demonstrating that our approach significantly improves the performance on detecting and reducing the spread of fake news and misinformation on the Web. Specifically, we showcase that our linguistic model, with the appropriate threshold, is able to achieve classification accuracies that exceed 99\%. To evaluate our approach, we have trained our model with the Fake News Corpus which includes 9 million articles labeled as fake and real. To the best of our knowledge, this is the biggest corpus in the research community.

\textbf{Roadmap.} The rest of this work is organized as follows. In Section 2 we describe related work from the literature. In Section 3 we describe our approach. In Section 4 we describe our experimental setup and detail the performance of our approach. In Section 5 we conclude this article.

\section{Related Work}

The task of fake news detection is similar to various other interesting challenges ranging from SPAM detection to rumor detection \cite{McCreadie2015}. In recent years, researchers are seeking to better define and characterize misinformation and its place in the larger information ecosystem \cite{Ruchansky:2017}. An important aspect of characterizing misinformation is to understand how people perceive the credibility of information. People usually tend to believe news that confirm what they already know, or what they already believe to be true \cite{Vosoughi1146}. News that go contrary to their beliefs (no matter how true the news are), may be met with high degrees of resistance. Thus, presenting people with the facts does not necessarily change their minds - several people keep on believing the fake news. To make matters worse, repeating the fake news, even in the context of refuting them, just makes them stronger. Thus, it seems that we need to explore non-obvious approaches to fight misinformation \cite{Lazer1094}.

Facebook is already partnering with fact-checking organizations. Facebook users are able to flag articles they suspect contain false information. These articles are then handed over to an independent evaluation centre. When a false story is identified, rather than being removed, it is tagged with a warning that it contains fake news and appears lower down in users feeds. Recently, Facebook will provide to social scientists unprecedented access to its data so that they can investigate how the spread of fake news on social media influences elections\footnote{\url{https://www.nature.com/articles/d41586-019-01447-5}}. Another initiative aiming to help citizens make informed choices ahead of the 2017 French election is the First Draft News project CrossCheck, a collaborative verification programme involving technology firms including Facebook and Google. The project sees journalists from across France working together to find and verify online content, including photos, videos, memes, comment threads and news sites. Similarly, Washington Post asked its readers to use the term ''Fake News'' so as to report the fake news. However, this term was used maliciously and it ended up being not so successful. Besides, some effort has also been done to detect fake news, including approaches that apply text-based methods\cite{Conroy2015} and fact-checking through knowledge graphs\footnote{\url{http://www.breitbart.com/big-government/2017/08/23/delingpole-all-of-recent-u-s-warming-has-been-faked-by-noaa}}. 
However, the current fact-checkers and crowdsourcing initiatives have limitations since they cannot cope with the high volume of misinformation generated online, and are usually disconnected from the Web browser, which is the medium used from users to read and share misinformation.

A few early studies tried to detect fake news based on linguistic features extracted from the text of news stories \cite{shu2018b},\cite{shu2018},\cite{Zhang:2018}. Recent studies have also shown that social networking features play a very important role in detecting fake news \cite{Vosoughi1146}. Deep neural networks have been successfully applied to  fake news detection \cite{Ruchansky:2017},\cite{shu2018b},\cite{shu2018}. Technical details regarding these approaches are presented in the evaluation section. However, all the existing approaches are trying to solve the problem using only one signal of information (i.e., fact-checking web sites, linguistic features, social networking features). Most current studies on misinformation either focus on analysing the influence of the topology of the social network on the consumption and sharing of misinformation, or taking into account the linguistic characteristics. Also, most systems tend to focus on the technical and not on the human aspects of the problem (i.e., the motivations of the users when generating and spreading misinformation). Our model is inspired from the SPAM detection research. Our system will assemble all sources of signal, and will combine them into one signal score. The score will reflect how confident we are that the story is fake (or not), and explore relationships among news comments$’$ topicality, temporality, sentiment, virality and quality.


\section{Check-It System}

\begin{figure*}
  \includegraphics[width=18cm,height=6.0cm]{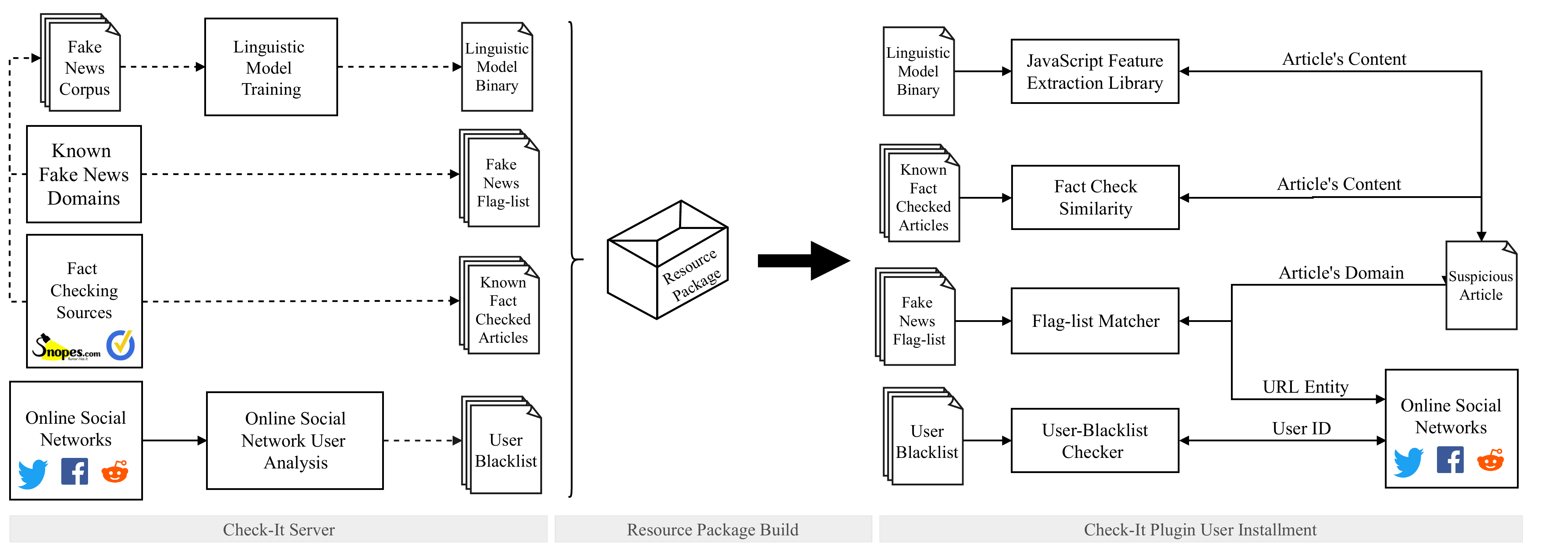}
  \centering
  \captionof{figure}{Architectural diagram for the Check-It System.}
\label{fig:check-it-system}
\end{figure*}

Check-It satisfies a series of user-centric functional requirements revolving around the user's data privacy, as listed below:

\begin{itemize}
    \item \textbf{Preserve User Privacy}: Check-It plugin should work locally, on the user's web browser, without the need of external communication (i.e. a RESTful API).
    \item \textbf{Highly Confident Identification}: Check-It labels a piece of news as fake if it is highly confident about it.
    \item \textbf{Low Response Time}: All the required resources, such as the flag-list and linguistic model, are efficiently loaded in the user's web browser. Also, the interconnected components of the plugin have been developed so as to have low response time.
    \item \textbf{Lightweight Computation}: Asynchronous processing and parallelization is taken place so as to minimize the load of the plugin.
\end{itemize}
Thus, our main objective is: \emph{to provide a Web browser plugin that detects efficiently and timely the fake news articles respecting the user's privacy.}

As depicted in Figure \ref{fig:check-it-system}, Check-It  system consists of four main components that function as a pipeline for fake news identification on the web. The Flag-list Matcher component matches domains of news articles to Known Fake News Domains and Fact Checks; the Fact Check Similarity component compares a piece of news against Known Fact Checked Articles labelled as fake from Fact Checking organizations, such as Politifact\footnote{\url{https://www.politifact.com/}} and Snopes\footnote{\url{https://www.snopes.com/}}; the Online Social Network User Analysis component is responsible for analyzing user behavior in social networks and producing a User-Blacklist of fake news propagators; and lastly, the Linguistic Model component, an artificial intelligence model, has been trained on linguistic features, of the Fake News Corpus, for the detection of fake news articles. 

Check-It preserves the user's privacy, whilst providing the appropriate functionality and performance, by loading the required resources locally, on the user's device. These resources are combined in a Resource Package, which includes the Fake News Flag-lists, the Known Fact Checked Articles, the User-Blacklist and the binary-produced Linguistic Model. The Resource Package is available by the Check-It Server. The only communication between the Check-It Server and the user, is during the installation of the plugin, where the required resources are downloaded and installed on the user's end (user-blacklist, fake news flag-list, known fact checked articles, and linguistic model), and any critical updates on those resources.

At the Check-It Plugin User Installment, the resources are loaded within the plugin, and assigned to their respective components. In addition to the Fact Check Similarity, Flag-list Matcher and User-Blacklist Checker, the Linguistic Model requires the features from the article's to be extracted. To this end, the JavaScript Feature Extraction Library was developed, responsible to capture the required features from within the article, and use them as input to the Linguistic Model Binary.


\subsection{Flag-list Matcher}

Some domain names are well known for spreading misinformation. Whether they do it on purpose, or for fun (such as satire), the information they provide is frequently not accurate and they should not be used as trusted sources of news. Currently there are several lists (which we call them \emph{flag-lists}) with contain domain names that host sites known for spreading misinformation. These lists are typically established and maintained by researchers or volunteers whose aim is to warn Internet users by ``flagging'' information sources of dubious credibility. The ``flagging'' provides some explanation with respect to why a domain name is included in the flag-list. For example, the flag may be ``fake news,'' which means that the site spreads misinformation, or ``biased,'' which means that the site is known to promote a biased point of view. A non-exhaustive list of the flag-lists we have used includes Kaggle\footnote{\url{https://www.kaggle.com/mrisdal/fake-news}}, OpenSources\footnote{\url{https://raw.githubusercontent.com/BigMcLargeHuge/opensources/master/sources/sources.csv}}, Greek-Hoaxes\footnote{\url{https://raw.githubusercontent.com/Ellinika-Hoaxes/Greek-Hoaxes-Detector/master/data/data.json}}, and several others. 

Our system has been designed in order to be easily configurable with respect to the flag lists it takes into account. URL flag-lists and domain name checking is the simplest way for an initial, fast assessment of the trustworthiness of a news article. Unfortunately, flag-lists do not test the truthfulness of the article itself: they just comment on the reputation of the website publishing the article. In that respect, flag-lists can be very helpful as long as they identify sites that consistently engage in disinformation campaigns or in propaganda spreading, in which case they can easily flag articles hosted in dubious sites. Nevertheless, one might want to be able to reason about the credibility of articles hosted in dubious web sites. To further assess the validity of such articles we use (i) fact-checking web sites (section 3.2) and (ii) machine learning approaches (sections 3.3 and 3.4), as we describe below. 

\subsection{Fact Check Similarity}

A number of Web sites currently in operation are dedicated to combating propaganda, misinformation, and hoaxes circulating on the Internet. These sites typically employ professional journalists (or even volunteers) who invest the time to research and comment on the truthfulness of articles shared on the web and on online social media, focusing primarily on evaluating the accuracy of articles or stories that become viral and receive increased user attention \cite{thorne-vlachos-2018-automated}. Once the truthfulness or falsehood of an article is established, these web sites publicize their findings and associated information (URL etc). Check-It capitalizes on fact-checking web sites, by cross checking every article processed by its plugin against a list of fact-checking web sites, generating an informative warning when an article happens to be found listed on these web sites. 

\subsection{Online Social Network User Analysis}


Since OSNs play an important role in the propagation of fake news \cite{Lazer1094}, we have incorporated another signal in the Check-It toolkit. The idea behind the OSN signal is to provide a \emph{dynamic user-blacklist}, matching user IDs with a falsity score, indicating the likelihood of a user to post fake news articles. The user-blacklist is dynamically generated by continuously processing OSN data and applying a DeGroot-based user probabilistic model \cite{DeGroot} for the user falsity score calculation. DeGroot model is used since it introduces a simple mechanism of opinion propagation: every individual forms her opinion by averaging her own opinion with those of her friends. The process is repeated until all opinions converge. Although the mechanism is simple, it models sufficiently opinion diffusion and incorporates elaborate characteristics of the process \cite{DeGroot}. Figure \ref{fig:osn-signal-architectural-diagram} presents the overall pipeline of the module and its components, which we describe in the next paragraphs. 

\noindent
\begin{minipage}{\linewidth}
\includegraphics[width=\linewidth, keepaspectratio]{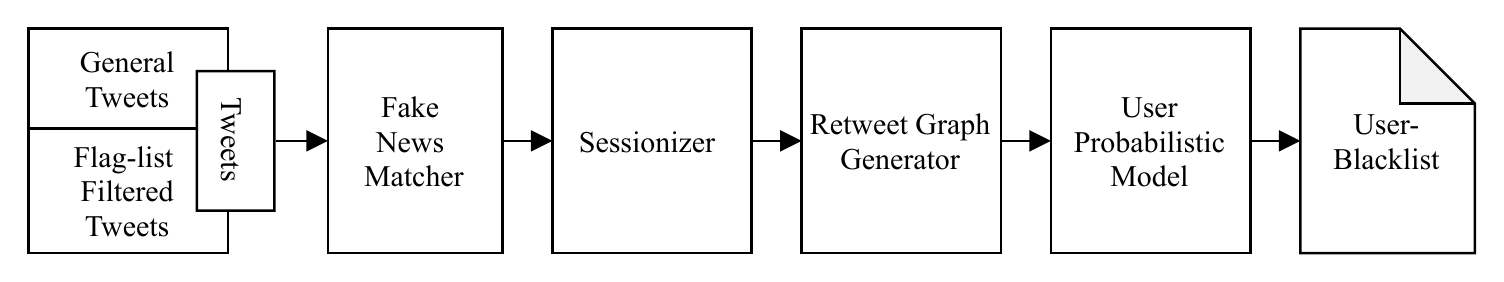}
\captionof{figure}{Architectural diagram for the social network signal.}
\label{fig:osn-signal-architectural-diagram}
\end{minipage}

The system design of Check-It facilitates integration with multiple OSN platforms. Currently, we only support Twitter due its massive popularity and the ease-of-access to its data stream via the Twitter Streaming API\footnote{\url{https://developer.twitter.com/en/docs/tweets/filter-realtime/overview}}. In particular, our system consumes tweets from two sources: a) tweets from the general public and b) tweets containing URLs of known fake news domains. The output of the system is a \emph{User-Blacklist} of fake news propagators.

The Flag-list Matcher component is responsible to mark tweets that contain a URL entity and positively answer the following question: \emph{Does the URL originate from a suspicious domain?} The tweets that have not been marked by the Flag-list Matcher are ordered in a timely manner and processed by the session-based model in groups of 1-hour sessions (Sessionizer task). A similar approach has been used in \cite{Vosoughi1146}. Then, each session is assigned to the Retweet Graph Generator, which is responsible for the creation of the retweet graph of the session. A retweet graph $G=(V, E)$ consists of nodes $u,v \in V$ depicting users and edges $(u,v) \in E$ representing the retweet action between users $u$ and $v$.  After the generation of the retweet graph, the User Probabilistic Model is applied in order to calculate the falsity score per user and produce the User-Blacklist. Initially, each user $u_i$ is assigned with a falsity score of $p_i^{(0)}=0$. Next, we briefly present the user probabilistic model that is based on the DeGroot's Learning Model.  

Let \emph{A} be the adjacency matrix of the retweet graph \emph{G}. We have that \emph{A(u,v)=1} if \emph{u} retweeted \emph{v}. We create a transition matrix \emph{T} by inverting the edges in \emph{A} (as the influence flows from the retweeted user to the user who retweeted him or her), adding a self-loop to each of the nodes and then normalizing each row in \emph{A} so it sums to 1 (meaning that each user is equally influenced by every user he or she retweets). Matrix \emph{T} includes the weight a node adds on another based on the sharing of fake news articles. We then associate a falsity score $p_i^{(0)}=1$ to every user who posted a suspicious tweet and $p_i^{(0)}=0$ to all who did not. Lastly, we create new scores $p(t)$ using the updating rule $p(t)=T \cdot p(t-1)$. In summary, the falsity score of a user increases if that user posts or retweets a suspicious tweet (a tweet that contains a URL from the flag-list). 

\subsection{Linguistic Model}
\label{sub:linguistic-model}

The linguistic component analyzes the actual content of the news article. Check-It extracts from an article's headline and body specific linguistic features, which have been widely used to detect fake news \cite{HorneA17, rubin-etal-2016-fake, Wang2018FiveSO}. These features are  used as input to a Deep Neural Network (DNN), which has been trained to predict the article's veracity. Rather than traditional machine learning, the deep learning approach was used due to the performance amplification it can achieve in the detection of fake news \cite{Ruchansky:2017, EANN2018} as well as in other problems addressed with artificial intelligence techniques. Next, we present an overview of the article dataset, the different linguistic features, and the DNN model.

\subsubsection{Dataset Overview}
\label{sec:fake-news-corpus}

Online news articles can be collected from different sources, such as news agency homepages, search engines, and social media websites. However, the manual determination of the veracity of news is a challenging task, usually requiring annotators with domain expertise. Check-It makes use of Fake News Corpus\footnote{\url{https://github.com/several27/FakeNewsCorpus}}, an open source dataset composed of \textbf{9 million news articles}. These articles originate from a curated list of 1001 domains collected from opensources.co. The entries are divided into 12 groups: \emph{fake news}, \emph{satire}, \emph{extreme bias}, \emph{conspiracy theory}, \emph{rumor mill}, \emph{state news}, \emph{junk science}, \emph{hate news}, \emph{clickbait}, \emph{political}, and \emph{credible}. In the scope of Check-It, we focus solely on the \emph{fake news} and \emph{credible} categories of the dataset, consisting of 1 million and 2 million articles respectively. As the dataset describes, fake news is considered when originating from ``\emph{sources that entirely fabricate information, disseminate deceptive content, or grossly distort actual news reports}'', whereas credible are ``\emph{sources that circulate news and information in a manner consistent with traditional and ethical practices in journalism}''.

\subsubsection{Linguistic Features}

Fake news detection on traditional news media mainly relies on news content, such as the headline and the body of an article. We compute different linguistic features that can be found in the headline and body of articles, in order to extract discriminative characteristics for the detection of fake news. These features are extracted and fed to the DNN model via the JavaScript Feature Extraction Library at Check-It plugin User Installment (Figure \ref{fig:check-it-system}. We group these features into 3 broad categories: \emph{stylistic}, \emph{complexity} and \emph{psychological}.

\textbf{Stylistic Features}: These are based on natural language processing to understand the syntax and text style of each article body and headline. Text style features include the frequency of stop-words,  punctuation,  quotes, negations and words that appear in all capital letters, whereas syntactical features include the frequency of Part-of-Speech tags in the text.

\textbf{Complexity Features}: These are based on deeper natural language processing computations aiming at capturing the overall intricacy of an article or headline. This intricacy can be computed based on several word-level metrics that include readability indexes and vocabulary richness. Specifically, we compute the Gunning Fog, SMOG Grade, and Flesh-Kincaid grade level readability indexes. Each measure computes a grade level reading score based on the number of complex words (e.g. over 3 syllables). A higher index means a document takes a higher education level to read. Moreover, we compute the Type-Token Ratio, which can be defined as the number of unique words divided by the total number of words in the article. In order to capture the vocabulary richness of the content, we also compute the number of hapax legomenon and dis legomenon, which correspond to phrase that occurs only once and twice within a context.

\textbf{Psychological Features}: The psychological features are based on the count of words found in expert dictionaries that are associated with different psychological processes. These dictionaries include the negative and positive opinion lexicon \cite{huliu}, and the moral foundation dictionary \cite{moralfoundation}. The sentiment score is computed via the AFINN sentiment lexicon \cite{afinn}, a list of English terms manually rated for valence. The AFINN sentiment score is defined as an integer number between -5 and +5, indicating the negative and positive score respectively. 

\subsubsection{Feature Selection}
The stylistic, complexity and psychological features are extracted from both the headline and body of the articles in the dataset, summing in 534 features. Such a large number of features results in an extensive model and deem the local execution as inadequate. In addition, unnecessary features can have side-effects during the model's training, decreasing training speed, model's interpretability, and generalization performance. In order to mitigate these issues, we proceed with a feature selection process to capture the 20 most descriptive features that facilitate the classification of news articles into fake or reliable. Below, we describe the feature selection process that is applied:

\begin{enumerate}
\item \textbf{Missing Values}: Remove features with a high percentage of missing values e.g. 60\%. Such features are not useful for the classification tasks as they do not carry any information, and can also affect the performance of the model.

\item \textbf{Single Unique Values}: Remove features with a single unique value, which have zero variance and have no contribution to the training of the model.

\item \textbf{Collinear Features}: Remove highly correlated features, which may lead to decreased generalization performance on the test set due to high variance and less model interpretability. These features are selected based on a specified correlation coefficient value (i.e., Pearson correlation coefficient).

\item \textbf{Zero Importance}: Calculate the importance of the remaining features according to a gradient boosting decision tree model, and remove features with zero importance. 

\item \textbf{Low Importance}: This step builds on the feature importance calculated in step (4), and its task is to remove features with low importance as they do not contribute to the total predefined importance. Principal Components Analysis (PCA) is used, keeping only the required principal components so as to retain a certain percentage of the variance (i.e, 95\%). 
\end{enumerate}

\noindent
The above feature selection process resulted in removing 134 features. From the remainder, the 20 most important were selected based on their importance scores, as extracted from step (4) (Table \ref{table:20-most-important-features}). These include the average number of stop-words in a sentence, the ratio of uppercase letters in the headline and the AFINN sentiment score.

\begin{table}[h!] 
\begin{center}
\begin{tabular}{|p{0.05\linewidth}|p{0.55\linewidth}|p{0.1\linewidth}|p{0.15\linewidth}|}
\hline
\textbf{No.} & \textbf{Feature} & \textbf{Score} & \textbf{Type} \\ \hline \hline
1 & Total number of lines & 0.0693 & Body \\ \hline
2 & Avg. number of stop-words per sentence & 0.0185 & Body \\ \hline
3 & Ratio of uppercase letters & 0.0177 & Headline \\ \hline
4 & Ratio of uppercase letters & 0.0152 & Body \\ \hline
5 & Avg. number of uppercase words per sentence & 0.0142 & Headline \\ \hline
6 & Avg. number of characters per word & 0.0141 & Body \\ \hline
7 & Ratio of alphabetic letters & 0.0139 & Headline \\ \hline
8 & Number of proper nouns (NP) & 0.0128 & Body \\ \hline
9 & Avg. number of sentences beginning with lowercase letter & 0.0126 & Body \\ \hline
10 & Avg. AFINN sentiment score & 0.0123 & Body \\ \hline
11 & Total number of characters & 0.0122 & Headline \\ \hline
12 & Ratio of digits & 0.0122 & Body \\ \hline
13 & Avg. number of sentences beginning with uppercase letter & 0.0122 & Body \\ \hline
14 & Ratio of alphabetic letters & 0.0119 & Body \\ \hline
15 & Number of genitive markers (POS) & 0.0116 & Body \\ \hline
16 & Number of colon or ellipsis & 0.0116 & Headline \\ \hline
17 & Total number of words beginning with uppercase letter & 0.0113 & Body \\ \hline
18 & Number of colon or ellipsis & 0.0102 & Body \\ \hline
19 & Avg. number of characters per word & 0.0096 & Headline \\ \hline
20 & Avg. number of stop-words per sentence & 0.0094 & Headline \\ \hline
\end{tabular}
\end{center}
\caption{Table with the 20 most important features as resulted from the feature selection process.}
\label{table:20-most-important-features}
\end{table}

\subsubsection{Deep Neural Network Model}

Similar to the linguistic feature selection, the proposed DNN model is compliant to the functional requirements set at the beginning of the project. It is a prerequisite that the model is compatible with conventional user devices and modern web browsers, as it is available as a traditional web browser plugin. Additional requirements are the low response time, lightweight computations and high confidence for the output. In order to address these challenges, the proposed DNN model adopts the cone-like structure, referred to as the bottleneck principle, and is known to perform well with numerical features \cite{TishbyZ15, HeZRS15}. The structure of the model is depicted in Figure \ref{fig:dnn-architectural-diagram}.

Before feeding the data into the DNN model, any categorical data are transformed into numerical, either via discretization or one-hot encoding, depending on the particulars of the input. As a result, each data entry is represented as a vector of numerical features. After the pre-processing, the data is used as input to the DNN model via the model's input layer.  

The next layer is a Batch Normalization Layer \cite{batchnormalization} which is responsible for the normalization of the activations of the previous layer (input layer) at each batch. Neural networks work better when the input data have zero mean and unit variance, as this enables faster learning and higher overall accuracy. A Batch Normalization Layer can achieve this by transforming and maintaining the mean and variance of its input close to zero. Next, the normalized output enters a set of fully connected layers (dense layers) that form the bottleneck. Such a bottleneck has been shown to result in automatic construction of high-level features. In our implementation, we experimented with multiple architectures, settling in a sequence of 5 layers that consist of 512, 256, 128, 64 and 32 neurons respectively. The final sequence is the one that provided the best results in our task. The units of the network are activated using the hyperbolic tangent activation function (tanh) since it is a better fit when working with standardized numerical data. 

Finally, in the DNN model's classification layer, one neuron per class is used with the softmax activation function to produce the probability pair of $P_{real}$ and $P_{fake}$, which correspond to the probability of the article being real or fake respectively.

\noindent

\begin{minipage}{\linewidth}
\includegraphics[width=1.05\linewidth, keepaspectratio]{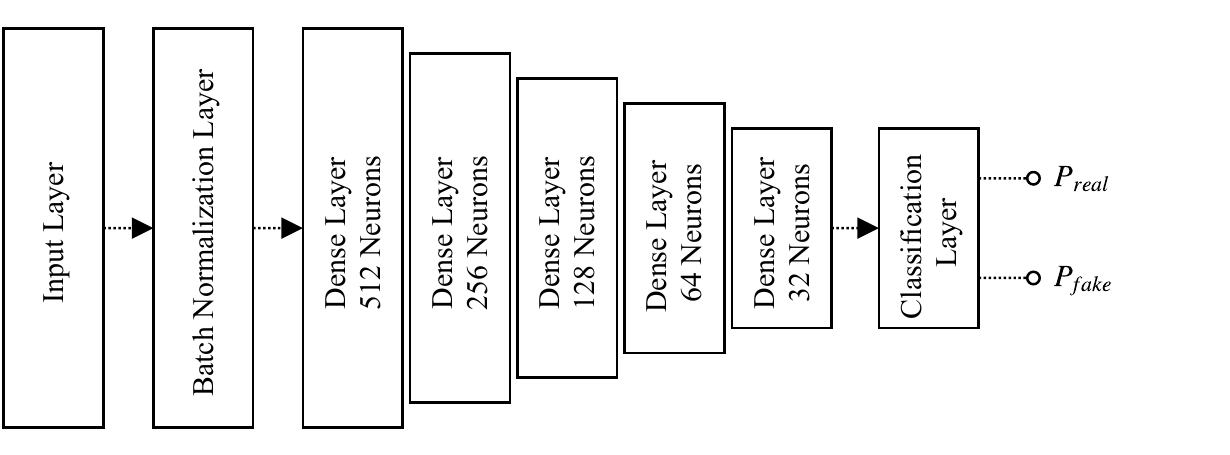}
\captionof{figure}{Architectural diagram for the deep neural network model used in the linguistic component.}
\label{fig:dnn-architectural-diagram}
\end{minipage}

\section{Evaluation}
For the evaluation of the Check-It plugin, we focus on the linguistic model and the user-blacklist generated by the Online Social Network User Analysis component. The Fake News Flag-lists and Known Fact Checked Articles are left out of the system evaluation since they provide us with 100\% accurate results. The task of these components is to transfer facts from the knowledge experts, such as the news site reviewers, from which the curated list of fake news domains was collected, and fact checking organizations consisting of journalists, reporters and experts from related fields. 

\subsection{Linguistic Model Evaluation}

For the implementation of the linguistic model presented in Section \ref{sub:linguistic-model}, Python Keras\footnote{\url{https://keras.io/}} has been used with Tensorflow\footnote{\url{https://www.tensorflow.org/}} as back-end. The training epochs for the model have been fixed at 100 with mini-batches of 128. During training, \emph{categorical cross-entropy} \cite{Goodfellow-et-al-2016} has been used as loss function and \emph{Adam} \cite{adamoptimizer} as the optimization function. Τo prevent the model from over-fitting, an \emph{early stopping} mechanism has been used. Early stopping is responsible for interrupting the training if the validation loss does not drop for 10 consecutive epochs. All the experiments were run in a stratified 3-fold cross validation, and executed on a Virtual Machine with Ubuntu 16.4, 16 VCPUs and 32GB of RAM. For the different parameters and the sake of training time, we also used Google's Colab\footnote{\url{https://colab.research.google.com/}}, a 12 hour free subscription to a Google Cloud VM with 13 GB of RAM and a Tesla K80 GPU. Finally, in order to be compatible with the user's internet browser, the model was exported with Tensorflow JS\footnote{\url{https://www.tensorflow.org/js}}. Tensorflow JS is a library for developing, training and exporting deep learning models in JavaScript, and deploying in the web browser.


We compared our system against 3 state-of-the-art works \cite{potthast2017, shu2018, shu2018b}. Note that for a fair comparison we chose baselines that only consider news contents, similar to our approach.The selected datasets include Buzzfeed News (\emph{BF}) and Politifact (\emph{PF}), which are publicly available in the authors Github\footnote{\url{https://github.com/KaiDMML/FakeNewsNet/tree/master/dataset}} repository. For evaluation metrics, we use accuracy, precision, recall and F1 score.

Shu et al. 2018 \cite{shu2018} utilize the \emph{BF} and \emph{PF} datasets in their work. The authors extracted news content features based on a combination of the vector space model and rhetorical structure theory (RST) \cite{rstfakenews} and the Linquistic Inquiry and Word Count (LIWC) lexicon \cite{liwc2015}, a widely used bundle of lexicons, that are able extract psycholignuistic features to capture deception within the articles. These features were used to train two separate SVM classifiers, namely $SVM_{RST}$ and $SVM_{LIWC}$. Furthermore, Shu et al. 2018b \cite{shu2018b}, utilize the \emph{GC} and \emph{PF} datasets to train several models, including an SVM, Logistic Regression (LR), Naive Bayes (NB) and a Convolutional Neural Network (CNN), focusing on one-hot vector representation of the data. Potthast et al. \cite{potthast2017} train 4 different Random Forest (RF) classifiers that consider the style and topic of the articles, 2 of them being generic, namely $GRF_{STYLE}$ and $GRF_{TOPIC}$, and 2 of them considering the political orientation of the articles, namely $ORF_{STYLE}$ and $ORF_{TOPIC}$. The authors utilize the \emph{BF} dataset, having information regarding the article's political orientation.

\begin{table}
    \begin{center}
    \begin{tabular}{| p{0.27\linewidth} | p{0.15\linewidth} | p{0.08\linewidth} | p{0.08\linewidth} | p{0.08\linewidth} | p{0.08\linewidth} |}
    \hline
    \textbf{Reference} & \textbf{Model} & \textbf{Acc.} & \textbf{P} & \textbf{R} & \textbf{F1} \\ \hline \hline
    
    \multirow{2}{*}{Shu et al. 2018 \cite{shu2018}}
            & $SVM_{LIWC}$ & 0.610 & 0.602 & 0.561 & 0.555 \\ \cline{2-6}
            & $SVM_{RST}$ & 0.655 & 0.683 & 0.628 & 0.623 \\ \hline
    
    \multirow{4}{*}{Potthast et al. \cite{potthast2017}}
            & $GRF_{STYLE}$ & 0.550 & 0.520 & 0.525 & 0.520 \\ \cline{2-6}
            & $GRF_{TOPIC}$ & 0.520 & 0.515 & 0.515 & 0.510 \\ \cline{2-6}
            & $ORF_{STYLE}$ & 0.550 & 0.535 & 0.540 & 0.535 \\ \cline{2-6}
            & $ORF_{TOPIC}$ & 0.580 & 0.555 & 0.555 & 0.560 \\ \hline
            
    \textbf{Check-It Model} & \textbf{$DNN$} & \textbf{\underline{0.703}} & \textbf{\underline{0.713}} & \textbf{\underline{0.703}} & \textbf{\underline{0.700}} \\ \hline
    \end{tabular}
    \end{center}
    \caption{Overall results on the comparison with the state-of-the-art for the Buzzfeed News (BF) dataset.}
    \label{table:overall-results-buzzfeed}
\end{table}

\raggedbottom

\begin{table}
    \begin{center}
    \begin{tabular}{| p{0.29\linewidth} | p{0.15\linewidth} | p{0.075\linewidth} | p{0.075\linewidth} | p{0.075\linewidth} | p{0.075\linewidth} |}
    \hline
    \textbf{Reference} & \textbf{Model} & \textbf{Acc.} & \textbf{P} & \textbf{R} & \textbf{F1} \\ \hline \hline
    
    \multirow{2}{*}{Shu et al. 2018 \cite{shu2018}}
            & $SVM_{RST}$ & 0.571 & 0.595 & 0.533 & 0.544 \\ \cline{2-6}
            & $SVM_{LIWC}$ & 0.637 & 0.621 & 0.667 & 0.615 \\ \hline
    
    \multirow{4}{*}{Shu et al. 2018b \cite{shu2018b}}
            & $SVM$ & 0.580 & 0.611 & 0.717 & 0.659 \\ \cline{2-6}
            & $LR$ & 0.642 & \textbf{\underline{0.757}} & 0.543 & 0.633 \\ \cline{2-6}
            & $NB$ & 0.617 & 0.674 & 0.630 & 0.651 \\ \cline{2-6}
            & $CNN$ & 0.629 & 0.807 & 0.456 & 0.583 \\ \hline
    \textbf{Check-It Model} & \textbf{$DNN$} & \textbf{\underline{0.722}} & \textbf{0.725} & \textbf{\underline{0.725}} & \textbf{\underline{0.722}} \\ \hline
    \end{tabular}
    \end{center}
    \caption{Overall results on the comparison with the state-of-the-art for the Politifact (PF) dataset.}
    \label{table:overall-results-politifact}
\end{table}

Next, we present the overall results of the state-of-the-art and compare them with the performance of our model. Table \ref{table:overall-results-buzzfeed} presents the results of the \emph{BF} dataset and Table  \ref{table:overall-results-politifact} presents the results of the \emph{PF} dataset. As displayed in Tables \ref{table:overall-results-buzzfeed} and \ref{table:overall-results-politifact}, Check-It linguistic model outperforms the state-of-the-art works. Our DNN, based on the deep learning paradigm, does not depend on handcrafted features, it rather generates abstract features, able to better capture the writing style of fake news \cite{varyingshades}. 






The datasets used for this experiment was to merely compare our model to the existing state-of-the-art models. Training our model with datasets of a few hundred records like the above, does not meet the expectations of deep learning \cite{Goodfellow-et-al-2016}. Thus, as described in Section \ref{sec:fake-news-corpus}, we trained on \emph{Fake News Corpus}, a dataset with millions of articles from domains, labelled as fake and real. \textbf{Our model is able to achieve an accuracy of 0.930, as well as 0.940 Precision, 0.937 Recall, and 0.937 F1 score}. 

\subsubsection{Optimization}

Despite the promising results, an error margin of 0.07 still exists.  
In order to reduce the error margin, we examined the number of false positives (FP) and true negatives (TN). 
Figure \ref{fig:dnn-thresholds} depicts the number of FP and TN as a function of the threshold, starting from 0.50 to 0.99 with step of 0.01. To achieve the maximum confidence, we chose the threshold to be 0.99, which resulted to 0 FP (Figure \ref{fig:dnn-confusion-matrix}). 
To test the generalization of our model and the performance with the adjusted threshold, an additional evaluation was made on several authoritative articles from news sources including ``The Guardian'', ``New York Times'', CNN and BBC. Specifically, 1158 articles were used as input to the DNN model with the adjusted threshold, from which only a single article was miss-classified as fake. 

\subsection{Online Social Network User Analysis Evaluation}
The task of Online Social Network User Analysis component is to build the User-Blacklist that includes the users that disseminate misinformation through social media. Our evaluation took place the time period from October $31_{st}$ 2018 to December $2_{nd}$ 2018, where we processed a total of 150 million tweets, from which 30 million tweets contained URLs from known fake news domains. In terms of user accounts, we processed a total of 8.1 million unique users. The users are categorized, based on their calculated falsity score, into users with low score [0-0.25), medium score [0.25-0.50), high score [0.50-0.75) and ultra high score [0.75-1.0). As we move from a low score to an ultra high score, the probability of a user, to disseminate a fake article, is increased. 

Taking into account the daily tweeting frequencies of the users in each group, we compared them with the frequencies of tweets containing URLs of known fake news domains (fake URLs). Figure \ref{fig:twitter-user-frequencies} depicts the frequencies of all the tweets and the tweets containing fake URLs. We see that users falling into the low falsity group have a large overall tweeting frequency with low frequency of tweets with fake URLs. Medium and high falsity groups present a rise on the frequency of tweets containing fake URLs. Users with ultra high falsity score seem to have lower overall frequency, but rather high frequency of tweets with fake URLs. Thus, the User-Blacklist consists of the users from the ultra high falsity group.

\begin{minipage}{\linewidth}
\includegraphics[width=\linewidth, keepaspectratio]{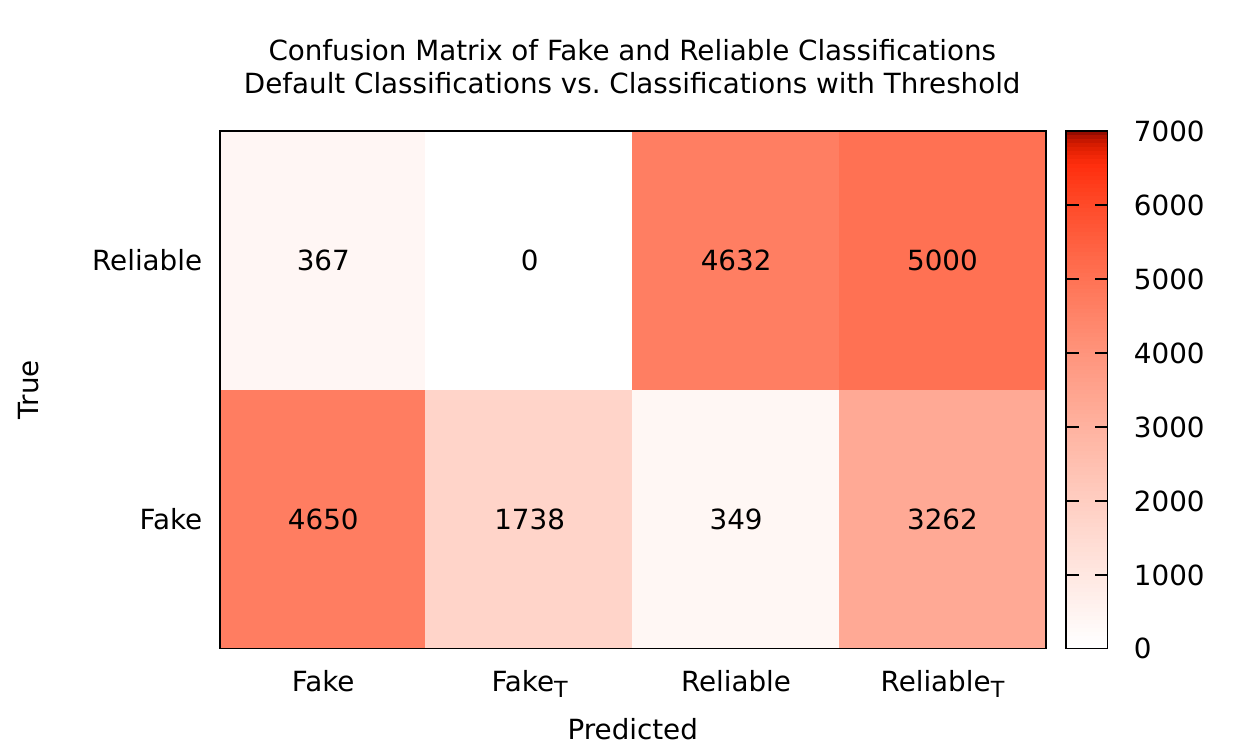}
\captionof{figure}{Confusion matrix of default classifications and classifications with threshold.}
\label{fig:dnn-confusion-matrix}
\end{minipage}

\begin{minipage}{\linewidth}
\includegraphics[width=\linewidth, keepaspectratio]{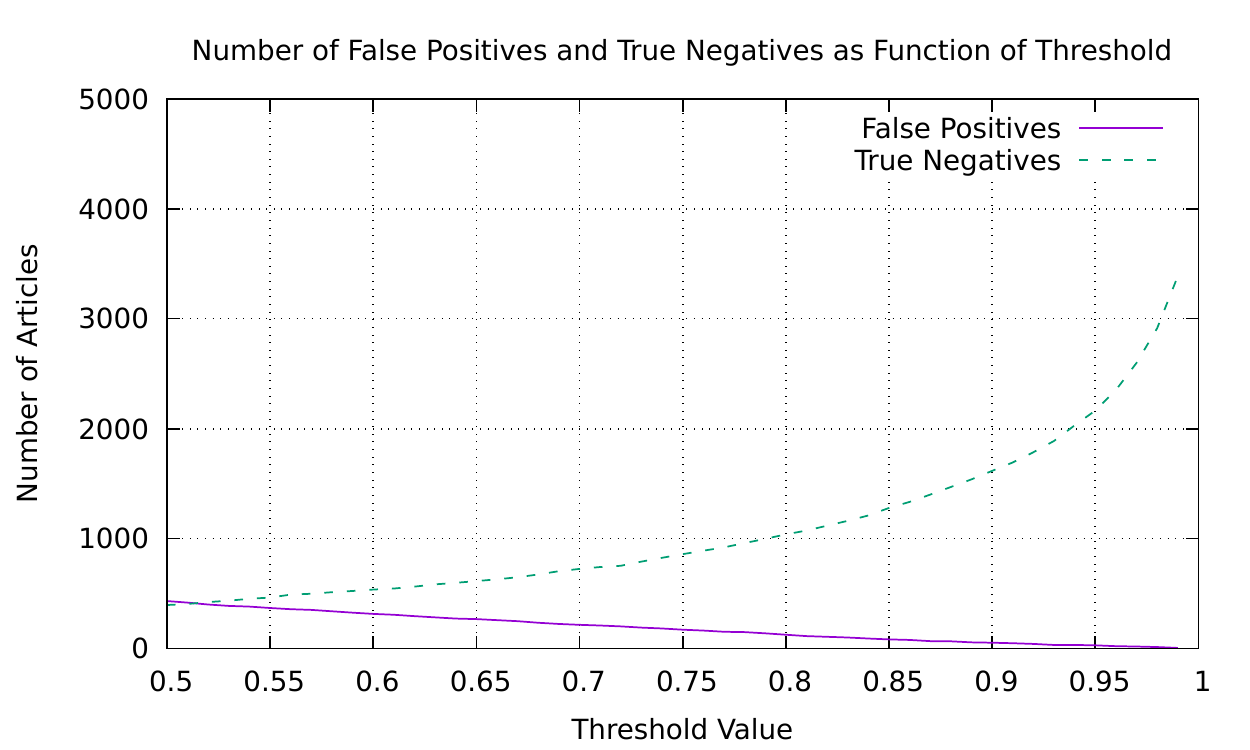}
\captionof{figure}{Number of False Positives and True Negatives as Function of Threshold}
\label{fig:dnn-thresholds}
\end{minipage}

\begin{minipage}{1\linewidth}
\includegraphics[width=1\linewidth, keepaspectratio]{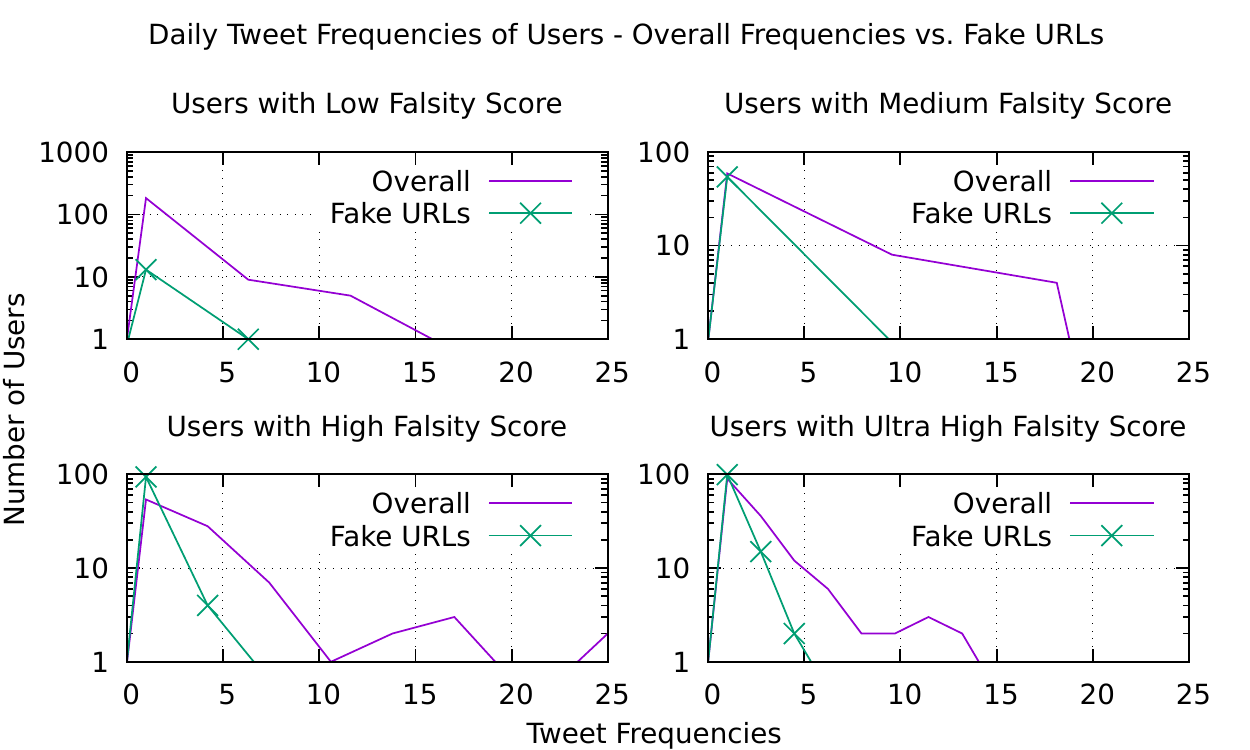}
\captionof{figure}{Tweeting frequencies of users that Check-It assigned ``Low'', ``Medium'', ``High'' and ``Ultra High'' falsity scores.}
\label{fig:twitter-user-frequencies}
\end{minipage}

\section{Conclusion}
In this paper, we presented Check-It, a fake news detection system, developed as a web browser plugin. Check-it aims to take a bold step towards detecting and reducing the spread of misinformation on the Web. To do so, it empowers its users with the tools they need to identify fake news. The major challenge of fake news detection stems from newly emerged news on which existing approaches only showed unsatisfactory performance. In order to address this issue, we propose a pipeline based on a variety of signals, ranging from domain name flag-lists to deep learning approaches. Extensive experiments showcase that Check-It is effective and can outperform the state-of-the-art models.


\begin{acks} 
This work is supported by the Google DNI grant (Check-it).
\end{acks}

\bibliographystyle{ACM-Reference-Format}
\bibliography{check-it.bib}

\appendix

\end{document}